\documentclass[12pt]{article}

\usepackage{float}
\usepackage{cite}
\usepackage{amsfonts}
\usepackage{amssymb}
\usepackage{amsmath}
\usepackage{xcolor}
\usepackage{mathtools}
\usepackage{mdwlist}
\usepackage{enumerate}
\usepackage{titlesec}

\usepackage{graphicx}

\def\gtwid{\mathrel{\raise.3ex\hbox{$>$\kern-.75em\lower1ex\hbox{$\sim$}}}}
\def\ltwid{\mathrel{\raise.3ex\hbox{$<$\kern-.75em\lower1ex\hbox{$\sim$}}}}
\def\square{\kern1pt\vbox{\hrule height 1.2pt\hbox{\vrule width 1.2pt\hskip 3pt
   \vbox{\vskip 6pt}\hskip 3pt\vrule width 0.6pt}\hrule height 0.6pt}\kern1pt}

\begin{document}

\begin{titlepage}

\begin{flushright}
UFIFT-QG-25-02
\end{flushright}

\vskip 2cm

\begin{center}
{\bf Recent Developments in Stochastic Inflation}
\end{center}

\vskip 1cm

\begin{center}
R. P. Woodard$^{\dagger}$
\end{center}

\vskip 0.5cm

\begin{center}
\it{Department of Physics, University of Florida,\\
Gainesville, FL 32611, UNITED STATES}
\end{center}

\vspace{0.5cm}

\begin{center}
ABSTRACT
\end{center}
This article is dedicated to the memory of Alexei Starobinsky. I begin
with some recollections of him and then review the generalization of his
wonderful stochastic formalism from scalar potential models to theories 
which interact with fermions and photons, and finally to theories with 
derivative interactions such as nonlinear sigma models and gravity. This
entails effective potentials generated by the usual field-dependent 
masses, as well as by field-dependent field strengths, and by 
field-dependent Hubble parameters. I also discuss secular loop 
corrections which cannot be captured by stochastic techniques.

\begin{flushleft}
\end{flushleft}

\vskip 2cm

\begin{flushleft}
$^{\dagger}$ e-mail: woodard@phys.ufl.edu
\end{flushleft}

\end{titlepage}

\section{Introduction}

Alexei Starobinsky was a genius whose contributions to cosmology include
one of the earliest models of primordial inflation \cite{Starobinsky:1980te}
which is still completely viable \cite{Tristram:2021tvh}. He also made the 
remarkable suggestion that quantum fluctuations from the epoch of primordial 
inflation might be observable today \cite{Starobinsky:1979ty} and can even
be used to constrain models of inflation \cite{Starobinsky:1982ee}. 

Of particular interest to me is the stochastic technique Alexei devised to
explain the secular growth one finds in the correlation functions of some
quantum field theories on inflationary backgrounds \cite{Starobinsky:1986fx}.
With Yokoyama, he even developed a way of summing up the leading secular 
corrections to all orders \cite{Starobinsky:1994bd}. I believe this technique
will eventually be recognized as equal in significance to the development of
the renormalization group by the late Ken Wilson for flat space quantum field 
theory and statistical mechanics \cite{Wilson:1971bg,Wilson:1971dh}. Wilson
won the 1982 Nobel Prize for this work and I feel that Alexei's contributions
would have been similarly honored, had he only lived.

This article consists of 7 sections. In the next one I share some personal
recollections of Alexei from the two decades during which I was privileged to
have known him. In section 3 I review the large secular loop corrections which
occur in scalar potential models on inflationary backgrounds, and I prove that 
Alexei's stochastic formalism recovers the leading logarithms produced at each 
order in perturbation theory. Section 4 explains how to handle charged scalars
and those with Yukawa couplings. In section 5 I discuss derivative interactions
involving scalars and gravity. Theories with such derivative interactions can 
experience another sort of leading logarithm loop correction which cannot be
described stochastically but is instead amenable to treatment by a variant of
the renormalization group. This is described in section 6. My conclusions 
comprise section 7.

\section{Remembering Alexei Starobinsky}

American elites are infamous for their view that the only interesting parts 
of my country are portions of the coasts, extending roughly from Boston to 
Washington in the East, and from Seattle to Los Angeles in the West. Many 
dismiss the rest of the nation as ``fly-over country'' which they only see 
while traveling between the parts they consider civilized. I'm told that 
Russian elites take a similar view of Moscow and Saint Petersburg. Alexei 
was not like that. Although he had many friends and collaborators outside of 
Russia, and could surely have emigrated had he wished, he stood by his 
country during hard times, and supported physics throughout its vast 
expanse. Indeed, he contracted the illness which took his life while 
attending an event in Kazan. And I first met Alexei in 2002, at a conference
in the Siberian city of Tomsk.

At that time of our first meeting I had been applying the Schwinger-Keldysh 
formalism \cite{Schwinger:1960qe,Mahanthappa:1962ex,Bakshi:1962dv,
Bakshi:1963bn,Keldysh:1964ud}, with dimensional regularization on de Sitter 
background, to work out loop corrections in renormalizable theories of 
massless, minimally coupled scalars which interact with themselves 
\cite{Onemli:2002hr}, or with electromagnetism \cite{Prokopec:2002jn,
Prokopec:2002uw}. It had long been known that the coincidence limit of the
massless, minimally coupled scalar propagator on de Sitter shows secular
growth \cite{Vilenkin:1982wt,Linde:1982uu,Starobinsky:1982ee}. This 
observation had eventually led to a full-blown proof that there are no
normalizable de Sitter invariant states in this theory \cite{Allen:1987tz}.
What I was showing is that the secular growth persists when realistic
interactions are added.

I confess to having felt a bit persecuted at the time of the Tomsk meeting
because both quantum field theory experts and cosmologists were, for 
different reasons, dismissing my carefully derived results as nonsense. 
The point which bothered quantum field theory people is that the secular 
growth I found breaks de Sitter invariance. They were convinced that de 
Sitter invariance should prove as powerful an organizing principle for QFT 
on de Sitter background as Poincar\'e invariance does on flat space 
background. They therefore suspected that my de Sitter breaking effects 
must derive from a poorly chosen vacuum, or from having posed unphysical 
questions \cite{Page:2012fn}, or from outright mistakes. On the other 
hand, some cosmologists altogether rejected the application of QFT on 
inflationary backgrounds. They argued that QFT expectation values amounted
to taking unphysical averages over portions of the wave function which had 
de-cohered and could not be simultaneously accessible to any real observer. 
Therefore, they concluded that my results were as unphysical as averaging 
over the alive-dead cat superposition in the famous paradox of 
Schr{\"o}dinger \cite{Abramo:2001dc}.

I spoke on my work and ended by asking, with some asperity, ``What is 
{\it wrong} with this?'' Alexei raised his hand and said that he didn't
think anything was wrong with my results, and that the secular growth they
contained could be predicted by his stochastic formalism. I was dumbfounded
(pleasantly) by a reaction so different from the ones I usually encountered. 
But I was also suspicious that there could be a trivial way of capturing the 
most important parts of dimensionally regulated and fully renormalized QFT 
calculations which required weeks of hard work. However, the possibility was 
too exciting to ignore, and I agreed to check Alexei's predictions against 
some three loop results I had not published. I screwed up the check at 
first, but Alexei patiently explained how to do it correctly, at which point 
I became convinced that his formalism must be right. Alexei was used to the 
skepticism of QFT experts that an ultraviolet finite, stochastic random 
variable could reproduce important features of an ultraviolet divergent, 
quantum field theory, and he must of sensed in me a willing tool to clarify 
for them a truth which he could see directly. In any case, he asked me to 
devise a proof from quantum field theory.

Alexei's friends and collaborators will confirm that he sometimes posed 
difficult problems! That particular one took me and Nick Tsamis three 
years to solve. When we finally found a proof \cite{Tsamis:2005hd} it
was my great pleasure to present it to Alexei during a conference at the
Lebedev Institute. Since then Alexei and I have overlapped at many sites:
Dubna, Moscow, Paris, Ann Arbor, Utrecht, Cargese, Tallinn, as well as 
innumerable virtual conferences. I treasured these opportunities to learn 
from him, and I was seldom disappointed. 

Like many brilliant scientists, Alexei was sometimes impatient with lesser 
men. It was my good fortune that he was always willing to talk with me. We 
twice began working together on projects, but never brought anything to 
completion. It saddens me that the opportunity to do so has been forever
foreclosed by his untimely passing.

Let me close this section by commenting that, for someone so renowned, 
Alexei was quite shy. My wife, Shun-Pei, who is also a shy person, picked 
up on this immediately when she first met him, during the fall of 2006.
She had recently discovered the first large logarithm induced by a loop of
gravitons in a dimensionally regulated and fully renormalized computation
\cite{Miao:2005am,Miao:2006gj}, and was attending a 3-month advanced 
school in Paris while still a graduate student. She and Alexei used to chat 
quietly in the corner during coffee breaks, while more gregarious and less 
accomplished physicists held forth loudly in the center. Years later, as a 
full professor in her own right, Shun-Pei organized the Taiwan portion of 
a Russia-Taiwan funding proposal with Alexei and Andrei Barvinsky. It was 
a good proposal and I'm sure it would have won support, however, the entire 
program was peremptorily terminated due to the worsening of relations that 
followed the Russian invasion of Ukraine, which Alexei opposed.

\section{Why Starobinsky's Formalism Works}

One can characterize the geometry of cosmology by its scale factor
$a(t)$, Hubble parameter $H(t)$ and first slow roll parameter $\epsilon(t)$,
\begin{equation}
ds^2 = -dt^2 + a^2(t) d\vec{x} \!\cdot\! d\vec{x} \qquad , \qquad H(t)
\equiv \tfrac{\dot{a}}{a} \qquad , \qquad \epsilon(t) \equiv -
\tfrac{\dot{H}}{H^2} \; . \label{geometry}
\end{equation}
Spacetime expansion means that the first derivative of $a(t)$ is positive,
which corresponds to $H(t) > 0$. Inflation means that the second derivative 
is also positive, which corresponds to $0 \leq \epsilon(t) < 1$. We know 
that this can occur because it's taking place now \cite{DES:2021wwk,
Kamionkowski:2022pkx}. The simplest interpretation of cosmological data 
\cite{Planck:2018vyg} is that it also took place during the very early
universe \cite{Geshnizjani:2011dk}, and Alexei's model 
\cite{Starobinsky:1980te} offers one of the best explanations for what 
caused it.

For light fields which are not conformally invariant the accelerated 
expansion of inflation can rip virtual, long wavelength quanta out of the 
vacuum so that they become real particles. This occurs maximally for 
massless, minimally coupled scalars, and for gravitons whose linearized 
equation of motion is the same \cite{Lifshitz:1945du}. On de Sitter 
background ($a(t) = e^{H t}$ with $H$ constant) the occupation number 
for a single wave vector $\vec{k}$ experiences staggering growth,
\begin{equation}
N(t,k) = [\tfrac{a(t) H}{2 k}]^2 \; . \label{occupation}
\end{equation}
This is what caused the (still unresolved) tensor power spectrum 
\cite{Starobinsky:1979ty} and its scalar cousin \cite{Mukhanov:1981xt},
which has been measured to three significant figures.

The exponentially increasing occupation numbers (\ref{occupation}) for 
super-horizon modes is what makes the coincident scalar propagator
grow \cite{Vilenkin:1982wt,Linde:1982uu,Starobinsky:1982ee}. The thing
which got me in so much trouble was asserting that this growth must
persist when interactions are present. The simplest example is a scalar
potential model,
\begin{equation}
\mathcal{L} = -\tfrac12 \partial_{\mu} \phi \partial_{\nu} \phi g^{\mu\nu}
\sqrt{-g} - V(\phi) \sqrt{-g} \; . \label{SPotMod}
\end{equation}
Few would argue that the expectation value of the stress tensor is 
unphysical. For any reasonable vacuum on the cosmological background
(\ref{geometry}) it must take the perfect fluid form,
\begin{equation}
\langle \Omega \vert T_{\mu\nu}(t,\vec{x}) \vert \Omega \rangle =
[\rho(t) + p(t)] \delta^0_{~\mu} \delta^0_{~\nu} + p(t) g_{\mu\nu} \; .
\label{Tmunu}
\end{equation}
For the case of $V(\phi) = \frac1{4!} \lambda \phi^4$ on de Sitter background
the 1-loop result can be absorbed into a renormalization of the cosmological
constant but the 2-loop results (suitably renormalized and with appropriate
1-loop state corrections to absorb exponentially falling terms) show secular 
growth \cite{Onemli:2002hr,Onemli:2004mb,Kahya:2009sz},
\begin{eqnarray}
\rho(t) &\!\!\! = \!\!\!& \tfrac{\lambda H^4}{2^7 \pi^4} \!\times\! \ln^2[a(t)]
+ O(\lambda^2) \; , \qquad \label{rho} \\
p(t) &\!\!\! = \!\!\!& \tfrac{\lambda H^4}{2^7 \pi^4} \!\times\! \Bigl\{-
\ln^2[a(t)] - \tfrac23 \ln[a(t)] \Bigr\} + O(\lambda^2) \; . \qquad \label{p}
\end{eqnarray}

It is time for a crucial digression on notation. The $\lambda \ln^2[a(t)]$ 
corrections in (\ref{rho}-\ref{p}) are {\it leading logarithm}; the $\lambda
\ln[a(t)]$ contribution to (\ref{p}) is a {\it sub-leading logarithm}. For
order $\lambda^N$ corrections in a quartic potential the leading logarithm
contribution is $\ln^{2N}[a(t)]$, and any term with fewer factors of $\ln[a(t)]$
is sub-leading. For a general monomial potential $V \sim \lambda \phi^K$ the
leading logarithm contribution at order $\lambda^{2N}$ is $\ln^{K\cdot N}[a(t)]$
\cite{Tsamis:2005hd}.

Vakif Onemli and I derived (\ref{rho}-\ref{p}) the hard way, by explicitly 
computing the various diagrams in dimensional regularization, then adding
counterterms and taking the unregulated limit. It took weeks. Starobinsky 
devised a way of trivially getting the leading logarithm result, not just for 
the stress tensor of a quartic potential but for any correlator in any scalar 
potential model (\ref{SPotMod}), and quite a bit before us 
\cite{Starobinsky:1986fx}. The technique is to replace the exact Heisenberg 
equation for the ultraviolet divergent quantum field $\phi(t,\vec{x})$,
\begin{equation}
-\tfrac1{\sqrt{-g}} \partial_{\mu} [ \sqrt{-g} \, g^{\mu\nu} \partial_{\nu} \phi] 
= -V'(\phi) \; , \label{Heisenberg}
\end{equation}
with a Langevin equation for the ultraviolet finite stochastic field 
$\varphi(t,\vec{x})$,
\begin{equation}
3 H [\dot{\varphi} - \dot{\varphi}_0] = -V'(\varphi) \; . \label{Langevin}
\end{equation}
The stochastic jitter $\dot{\varphi}_0$ is the time derivative of the infrared
truncated, free field mode sum,
\begin{equation}
\varphi_0(t,\vec{x}) = \int_{H}^{H a(t)} \!\!\!\!\!\!\!\!\!\! \tfrac{d^3k}{(2\pi)^3} 
\tfrac{H e^{i \vec{k} \cdot \vec{x}}}{\sqrt{2 k^3}} \Bigl\{ \alpha(\vec{k})
+ \alpha^{\dagger}(-\vec{k}) \Bigr\} \;\; , \;\; [\alpha(\vec{k}),
\alpha^{\dagger}(\vec{p})] = (2\pi)^3 \delta^3(\vec{k} \!-\! \vec{p}) \; . 
\label{jitter}
\end{equation}

There are two ways to use Starobinsky's Langevin equation (\ref{Langevin}). 
The first is to solve it perturbatively for $\varphi(t,\vec{x})$ as an expansion 
in powers of $\varphi_0(t,\vec{x})$. For the quartic potential this gives,
\begin{eqnarray}
\lefteqn{\varphi(t,\vec{x}) = \varphi_0(t,\vec{x}) - \tfrac{\lambda}{18 H} \!\!
\int_0^{t} \!\! dt' \, \varphi^3_0(t',\vec{x}) } \nonumber \\
& & \hspace{4cm} + \tfrac{\lambda^2}{108 H^2} \!\! \int_0^{t} \!\! dt' 
\varphi^2_0(t',\vec{x}) \!\! \int_0^{t'} \!\! dt'' \, \varphi^3_0(t'',\vec{x}) 
+ O(\lambda^3) \; . \qquad \label{Langsolution}
\end{eqnarray}
Substituting in the stress tensor and taking the expectation value gives the
leading logarithm expansion for the energy density (\ref{rho}) 
\cite{Tsamis:2005hd},
\begin{equation}
\rho(t) \longrightarrow \tfrac{\lambda H^4}{2^7 \pi^4} \ln^2[a(t)] \!\times\! 
\Bigl\{ 1 - \tfrac{\lambda}{12 \pi^2} \ln^2[a(t)] + \tfrac{53 \lambda^2}{2^4 3^4 
5 \pi^4} \ln^4[a(t)] + O\Bigl(\lambda^3 \ln^6[a(t)]\Bigr) \Bigr\} . \label{rhoexp}
\end{equation}
Note that I have carried this to {\it fourth order} in the loop expansion.

The second way of using Starobinsky's Langevin equation (\ref{Langevin}) is
to derive a Fokker-Planck equation for the time-dependent probability density
$\varrho(t,\varphi)$ \cite{Starobinsky:1994bd},
\begin{equation}
\dot{\varrho}(t,\varphi) = \tfrac{\partial}{\partial \varphi} \Bigl[ 
\tfrac{V'(\varphi)}{3 H} \varrho(t,\varphi)\Bigr] + \tfrac{H3}{8 \pi^2} 
\tfrac{\partial^2 \varrho(t,\varphi)}{\partial \varphi^2} \; . \label{FPEQN}
\end{equation}
If the potential is bounded below, and steep enough, the system will approach 
an equilibrium when the inward pressure of the classical force balances the 
outward pressure of inflationary particle production. In that case the left
hand side of equation (\ref{FPEQN}) vanishes and we can solve for the late
time probability density,
\begin{equation}
\varrho(t,\varphi) \longrightarrow N \exp\Bigl[-\tfrac{8\pi^2 V(\varphi)}{3 H^4}
\Bigr] \; . \label{latevarrho}
\end{equation}
This allows one to sum the series of leading logarithms. For example, the
late time limit of the energy density for the quartic potential is,
\begin{equation}
\rho(t) \longrightarrow \tfrac{3 H^4}{8 \pi^2} \tfrac{\Gamma(\frac34)}{
\Gamma(\frac14)} \; . \label{rholate}
\end{equation}
The physical interpretation is that inflationary particle production pushes
the scalar up its quartic potential until an equilibrium is reached with the
downward classical force.

Before giving the proof of Starobinsky's formalism let me comment on how 
very strange it seems from the perspective of quantum field theory. The 
original scalar potential model (\ref{SPotMod}) describes a quantum field 
which obeys a second order equation and whose commutator $[\phi(t,\vec{x}),
\phi(t',\vec{x}')]$ is nonzero for timelike and lightlike separations. The 
correlators of this quantum field are typically ultraviolet divergent and 
require both regularization and renormalization. In contrast, the infrared 
truncated mode sum (\ref{jitter}) for $\varphi_0(t,\vec{x})$ is a stochastic
random variable which commutes with itself everywhere, $[\varphi_0(t,\vec{x}),
\varphi_0(t',\vec{x}')] = 0$. Because the mode sum is cut off at $k = H a(t)$,
its correlators are ultraviolet finite. These same two properties are inherited
by the full field $\varphi(t,\vec{x})$, which obeys the first order equation
(\ref{Langevin}). Yet correlators of $\varphi$ agree with those of $\phi$ at
leading logarithm order. You can see why quantum field theory experts were
suspicious of Alexei's formalism!  

The first step in proving Starobinsky's formalism is to act the retarded Green's
function on the dimensionally regulated Heisenberg field equation 
(\ref{Heisenberg}) to derive the Yang-Feldman equation \cite{Tsamis:2005hd},
\begin{equation}
\phi(x) = \phi_0(x) - \int \!\! d^Dx' \sqrt{-g(x')} \, i \theta(t \!-\! t')
[\phi_0(x), \phi(x')] V'\Bigl( \phi(x')\Bigr) \; . \label{YFeqn}
\end{equation}
The free field $\phi_0(x)$ is,
\begin{equation}
\phi_0(t,\vec{x}) = \int_{H} \!\! \tfrac{d^{D-1}k}{(2\pi)^{D-1}} \Bigl\{
\alpha(\vec{k}) u(t,k) e^{i \vec{k} \cdot \vec{x}} + \alpha^{\dagger}(\vec{k})
u^*(t,k) e^{-i \vec{k} \cdot \vec{x}} \Bigr\} \; , \label{freefield}
\end{equation}
where $\alpha^{\dagger}$ and $\alpha$ are canonically normalized creation and
annihilation operators and the free field mode function in Bunch-Davies
vacuum \cite{Chernikov:1968zm,Schomblond:1976xc,Bunch:1978yq} is,
\begin{equation}
u(t,k) \equiv i \sqrt{ \tfrac{\pi}{4 H a^{D-1}} } \, H^{(1)}_{\frac{D-1}2}\Bigl(
\tfrac{k}{a H}\Bigr) \; . \label{modefunc}
\end{equation}
Note that I have expressed the retarded Green's function as $i \theta(t - t')$
times the commutator of two free fields. Note also that the infrared limit of
the mode function is,
\begin{equation}
k \ll H a \qquad \Longrightarrow \qquad u(t,k) = \tfrac{\Gamma(\frac{D-1}{2})}{
\sqrt{4 \pi H}} \Bigl( \tfrac{2 H}{k}\Bigr)^{\frac{D-1}2} \Bigl\{1 + O\Bigl(
\tfrac{k^2}{a^2 H^2}\Bigr) \Bigr\} \; . \label{IRlimit}
\end{equation}
This is why the free field mode sum (\ref{freefield}) has been infrared cut off 
at $k = H$ to exclude modes which are in correlated Bunch-Davies vacuum at the 
beginning of inflation \cite{Vilenkin:1982wt}. 

The Yang-Feldman equation (\ref{YFeqn}) is exact; iterating it would produce the
usual interaction picture expansion of the full field $\phi$ in powers of the
free field $\phi_0$. Factors of $\ln[a(t)] = Ht$ in correlators derive from two 
sources:
\begin{enumerate*}
\item{The integral of $u(t,k) u^*(t',k)$ from the infrared cutoff at $k = H$ to
the point $k = H \times {\rm min}[a(t),a(t')]$ at which the infrared limit 
(\ref{IRlimit}) breaks down; and}
\item{Vertex integrations over the retarded Green's function.}
\end{enumerate*}
The first source is associated with a pair of free fields which combine to give a 
propagator; the second source is also associated with the pair of free fields in 
the commutator of the retarded Green's function. Now recall that reaching leading 
logarithm order requires that each extra pair of vertices must contribute one 
factor of $\ln[a]$ for each pair of fields. In the interaction picture this 
translates to each pair of free fields (including those in the retarded Green's 
function) contributing a factor of $\ln[a]$. Because these factors derive from 
the two sources given above, correlators will be unaffected, at leading logarithms 
order, if we cut off the mode sum in (\ref{freefield}) at $k = H a(t)$ (at which 
point we can take $D=4$), and also retain only the term (\ref{IRlimit}) in the 
small $k/aH$ expansion of the mode function. 

Because the retarded Green's function involves a commutator, we must carry the
expansion for these terms out to include the first imaginary part,
\begin{equation}
D = 4 \qquad \Longrightarrow \qquad u(t,k) = \tfrac{H}{\sqrt{2 k^3}} \Bigl\{1 + 
\tfrac12 (\tfrac{k}{a H})^2 + \tfrac{i}{3} (\tfrac{k}{a H})^3 + \dots \Bigr\} \; . 
\label{modeexpand}
\end{equation}
Hence the retarded Green's function becomes,
\begin{eqnarray}
G(x;x') &\!\!\! = \!\!\!& i\theta(t \!-\! t') \!\! \int \!\! \tfrac{d^3k}{(2\pi)^3}
\, e^{i \vec{k} \cdot (\vec{x} - \vec{x}')} \Bigl\{ u(t,k) u^*(t',k) - u^*(t,k)
u(t',k)\Bigr\} \; , \qquad \\
&\!\!\! \longrightarrow \!\!\!& \tfrac{\theta(t - t')}{3 H} \times [\tfrac1{a^3(t')} 
- \tfrac1{a^3(t)}] \times \delta^3(\vec{x} \!-\! \vec{x}') \; . \qquad 
\label{retGreen}
\end{eqnarray}
Of course this allows us to perform the spatial integrations in the Yang-Feldman 
equation (\ref{YFeqn}). A final simplification becomes apparent after multiplying 
(\ref{retGreen}) by the measure factor $\sqrt{-g(x')} = a^3(t') = e^{3 H t'}$ in 
expression (\ref{YFeqn}). It follows that the factor of $1/a^3(t)$ in 
(\ref{retGreen}) can be neglected,
\begin{equation}
\int_0^{t} dt' a^3(t') \times [\tfrac1{a^3(t')} - \tfrac1{a^3(t)}] = t - 
\tfrac1{3 H} [ 1 - e^{-3 H t}] \; . \label{finalsimp}
\end{equation} 

The various truncations produce a completely different field $\varphi(t,\vec{x})$ 
whose correlators nonetheless agree with those of $\phi(t,\vec{x})$ at leading
logarithm order. The truncated Yang-Feldman equation obeyed by this field is,
\begin{equation}
\varphi(t,\vec{x}) = \varphi_0(t,\vec{x}) - \tfrac1{3 H} \!\! \int_{0}^{t}
\!\!\!\! dt' \, V\Bigl( \varphi(t',\vec{x}) \Bigr) \; . \label{truncLang}
\end{equation}
Here $\varphi_0(t,\vec{x})$ is precisely the infrared truncated free field
expansion given in (\ref{jitter}). Taking the time derivative of 
(\ref{truncLang}), and making some trivial rearrangements, gives Starobinsky's
Langevin equation (\ref{Langevin}). Note that we have not only established its
validity, we have clarified its precise relation to quantum field theory: {\it 
Starobinsky's formalism reproduces the leading logarithms of scalar potential 
models} (\ref{SPotMod}). 

\section{Passive Fields}

It is important to distinguish between {\it Active Fields}, which experience
inflationary particle production (\ref{occupation}), and {\it Passive Fields},
which do not. The prime examples of active fields are massless, minimally 
coupled scalars and gravitons; examples of passive fields include conformally 
coupled scalars, fermions and photons. We have seen that loops of active 
fields induce secular growth such as (\ref{rho}-\ref{p}). Passive fields do 
not induces such factors, but they can mediate them, and they can modify the 
way that active fields interact.

Passive fields typically contribute nonzero constants to correlators. For
example, one of several 1-loop electromagnetic contributions to the 
dimensionally regulated graviton self-energy on de Sitter background involves 
the coincidence limit of two field strengths \cite{Wang:2015eaa},
\begin{equation}
\langle \Omega \vert F_{\mu\nu}(x) F_{\rho\sigma}(x) \vert \Omega \rangle =
\tfrac{H^D}{(4 \pi)^{\frac{D}2}} \tfrac{\Gamma(D-1)}{\Gamma(\frac{D}2 + 1)}
\times [g_{\mu \rho} g_{\nu \sigma} - g_{\mu \sigma} g_{\nu \rho}] \; . 
\label{FF}
\end{equation}
Such contributions come from the full range of modes, from ultraviolet to 
infrared. They also involve the full passive field mode function, not just its 
infrared truncation. This is why no stochastic formulation of a passive field 
can be correct. For example, stochastically truncating electromagnetism would
render the field strength an ultraviolet finite, Hermitian operator whose
square must be positive. That is obviously inconsistent with the exact 
result (\ref{FF}) for the case of $\mu = \rho$ and $\nu = \sigma$.  

The right thing to do with passive fields is integrate them out in the presence
of a constant active field background. For passive fields coupled to an active
scalar field setting this scalar to a constant induces a mass for the passive
field, and integrating out the passive field results in a Coleman-Weinberg
potential \cite{Coleman:1973jx}. At this stage the active field has been reduced
to a scalar potential model on which Starobinsky's formalism can be applied.
Two theories for which this procedure has been applied are a real scalar which
is Yukawa-coupled to a fermion\cite{Miao:2006pn}, and a complex scalar which is 
coupled to electromagnetism \cite{Prokopec:2007ak}.

The Lagrangian for Yukawa theory is,
\begin{eqnarray}
\lefteqn{\mathcal{L} = -\tfrac12 \partial_{\mu} \phi \partial_{\nu} \phi 
g^{\mu\nu} \sqrt{-g} -\tfrac12 \delta \xi \phi^2 R \sqrt{-g} - V(\phi) \sqrt{-g} 
- f \phi \overline{\psi} \psi \sqrt{-g} } \nonumber \\
& & \hspace{6cm} + \overline{\psi} e^{\mu}_{~a} \gamma^a \Bigl(i \partial_{\mu}
- \tfrac12 A_{\mu bc} J^{bc} \Bigr) \psi \sqrt{-g} \; . \qquad \label{Yukawa}
\end{eqnarray}
The usual notation applies: $\gamma^a$ denotes the gamma matrices, the vierbein
is $e^{\mu}_{~a}(x)$, the spin connection is $A_{\mu bc}(x)$, and the spin
generator is $J^{bc}$,
\begin{equation}
g^{\mu\nu} = e^{\mu}_{~a} e^{\nu}_{~b} \eta^{ab} \quad , \quad 
A_{\mu bc} \equiv e^{\nu}_{~b} (e_{\nu c , \mu} - \Gamma^{\rho}_{~\mu\nu}
e_{\rho c}) \quad , \quad J^{bc} \equiv \tfrac{i}{4} [\gamma^{b}, 
\gamma^{c}] \; . \label{spinstuff}
\end{equation}
For constant $\phi$ the last term on the first line of (\ref{Yukawa}) can be
recognized as a fermion mass of $m = f \phi$. The massive fermion propagator on
de Sitter $i [\mbox{}_i S_{j}](x;x')$ was derived by Candelas and Raine 
\cite{Candelas:1975du}. Integrating out the fermion out of the scalar field
equation (with suitable renormalizations) gives,
\begin{eqnarray}
\lefteqn{\partial_{\mu} (\sqrt{-g} \, g^{\mu\nu} \partial_{\nu} \phi) - \delta 
\xi \phi R \sqrt{-g} - V'(\phi) \sqrt{-g} - f \overline{\psi} \psi \sqrt{-g} }
\nonumber \\
& & \hspace{0cm} \longrightarrow \partial_{\mu} (\sqrt{-g} \, g^{\mu\nu} 
\partial_{\nu} \phi) - [\delta \xi \phi R - V'(\phi)] \sqrt{-g} - f
i [\mbox{}_{i} S_{i}](x;x) \sqrt{-g} \; , \qquad \\
& & \hspace{0cm} \longrightarrow \partial_{\mu} (\sqrt{-g} \, g^{\mu\nu} 
\partial_{\nu} \phi) - [\alpha H^2 \phi + \beta \phi^3] \sqrt{-g} \nonumber \\
& & \hspace{3cm} + \tfrac{f H^3}{4 \pi^2} \tfrac{f \phi}{H} (1 + \tfrac{f^2 
\phi^2}{H^2}) \Bigl\{ \psi(1 + \tfrac{i f \phi}{H}) + \psi(1 - 
\tfrac{i f \phi}{H}) \Bigr\} \sqrt{-g} \; . \qquad \label{Yuksimp}
\end{eqnarray}
Here $\psi(x) \equiv \frac{d}{dx} \ln[ \Gamma(x)]$ is the digamma function and
the constants $\alpha \sim f^2$ and $\beta \sim f^4$ are arbitrary.

The Lagrangian of scalar quantum electrodynamics (SQED) is,
\begin{eqnarray}
\lefteqn{\mathcal{L} = - (D_{\mu} \phi)^* D_{\nu} \phi g^{\mu\nu} \sqrt{-g} 
- \delta \xi \phi^* \phi R \sqrt{-g} - V(\phi^* \phi) \sqrt{-g} } \nonumber \\
& & \hspace{8cm} - \tfrac14 F_{\mu\nu} F_{\rho\sigma} g^{\mu\rho} g^{\nu\sigma} 
\sqrt{-g} \; . \qquad \label{SQED}
\end{eqnarray}
Here $D_{\mu} \equiv \partial_{\mu} + i e A_{\mu}$ is the covariant derivative
operator. Setting the scalar equal to a constant endows the photon with a mass
$m^2 = e^2 \phi^* \phi$. The Lorentz gauge massive photon propagator on de 
Sitter background $i [\mbox{}_{\mu} \Delta_{\nu}](x;x')$ was derived in 
\cite{Tsamis:2006gj}. Using it to integrate the photon out of the scalar field
equation (with suitable renormalization) gives
,\footnote{The result was first
obtained by Allen as a mode sum \cite{Allen:1983dg}.}
\begin{eqnarray}
\lefteqn{ D_{\mu} (\sqrt{-g} \, g^{\mu\nu} D_{\nu} \phi) - \delta \xi \phi R 
\sqrt{-g} - \phi V'(\phi^* \phi) \sqrt{-g} } \nonumber \\
& & \hspace{0cm} \longrightarrow \partial_{\mu} (\sqrt{-g} \, g^{\mu\nu}
\partial_{\nu} \phi) - \phi [\delta \xi R \!-\! V'] \sqrt{-g} - e^2 \phi \,
i[\mbox{}_{\mu} \Delta_{\nu}](x;x) g^{\mu\nu} \sqrt{-g} \; , \qquad \\ 
& & \hspace{0cm} \longrightarrow \partial_{\mu} (\sqrt{-g} \, g^{\mu\nu}
\partial_{\nu} \phi) - \phi [\overline{\alpha} + \overline{\beta} \phi^* \phi]
\sqrt{-g} - \tfrac{3 e^2 H^2 \phi}{8 \pi^2} (1 + \tfrac{e^2 \phi^* \phi}{H^2})
\nonumber \\
& & \hspace{2cm} \times \Bigl\{ \psi\Bigl( \tfrac32 \!+\! \tfrac12 
[1 \!-\! \tfrac{8 e^2 \phi^* \phi}{H^2}]^{\frac12} \Bigr) + \psi\Bigl( \tfrac32 
\!-\! \tfrac12 [1 \!-\! \tfrac{8 e^2 \phi^* \phi}{H^2}]^{\frac12} \Bigr) \Bigr\} 
\sqrt{-g} \; . \qquad \label{SQEDsimp}
\end{eqnarray}
The constants $\overline{\alpha} \sim e^2$ and $\overline{\beta} \sim e^4$ are
again arbitrary.

The stochastic formulations of the scalar potential models (\ref{Yuksimp}) 
and (\ref{SQEDsimp}) agree with explicit dimensionally regulated and fully 
renormalized computations at 1-loop and 2-loop orders \cite{Miao:2006pn,
Prokopec:2007ak,Prokopec:2006ue,Prokopec:2008gw}. ``Leading logarithm'' in
these two theories means that each additional factor of $f^2$ in Yukawa, or
of $e^2$ in SQED, brings an extra factor of $\ln[a(t)]$. Note that ultraviolet 
regularization and renormalization are required to derive the Coleman-Weinberg 
potentials, however, the resulting scalar potential model is afterwards UV 
finite at leading logarithm order. The need for renormalization is reflected 
in the arbitrary finite parts associated with the constants $\alpha$ and 
$\beta$ for (\ref{Yuksimp}), and $\overline{\alpha}$ and $\overline{\beta}$ 
for (\ref{SQEDsimp}). 

Finally, it is worth commenting on signs. The Yukawa effective potential 
(\ref{Yuksimp}) is negative, so this system never reaches a stable equilibrium; 
the universe instead decays in a Big Rip singularity \cite{Miao:2006pn}. In 
contrast, the effective potential for SQED is bounded below, which means that 
the system approaches a stable equilibrium. However, it turns out that the 
vacuum energy is slightly reduced \cite{Prokopec:2007ak}, for essentially the 
same reason that a slab of dielectric is pulled into a charged, parallel plate 
capacitor. With the quartic potential result (\ref{rholate}) this means that
scalar models span the range of possibilities:
\begin{itemize*}
\item{The system may or may not approach a stable equilibrium; and}
\item{The vacuum energy may be increased or decreased.}
\end{itemize*}

\section{Differentiated Active Fields}

Because the secular growth factors of $\ln[a(t)]$ derive from logarithmic
integrations, the presence of a single derivative suppresses them. The
status of differentiated active fields is very similar to that of passive
fields: although they induce no secular loop corrections, they make important
contributions of order one which come as much from the ultraviolet as from
the infrared, and involve the full mode function. Hence it is also wrong to 
infrared truncate differentiated active fields. For example, in dimensional
regularization the coincidence limit of the doubly differentiated, massless, 
minimally coupled scalar propagator is \cite{Onemli:2002hr,Onemli:2004mb},
\begin{equation}
\lim_{x' \rightarrow x} \partial_{\mu} \partial'_{\nu} i\Delta(x;x') =
\langle \Omega \vert \partial_{\mu} \phi_0(x) \partial_{\nu} \phi(x) \vert
\Omega \rangle = -\tfrac{H^D}{2 (4\pi)^{\frac{D}{2}}} \tfrac{\Gamma(D)}{
\Gamma(\frac{D}2 + 1)} \times g_{\mu\nu}(x) \; . \label{ddprop}
\end{equation}
Were we to replace $\phi_0(x)$ by its infrared truncation $\varphi_0(x)$
the expectation value would be positive for $\mu = \nu$ because it is the 
square of a finite, Hermitian operator. However, the exact result (\ref{ddprop})
is {\it negative} for spatial $\mu = \nu$.

Rather than stochastically simplifying differentiated active fields, the
correct procedure is to integrate them out of the field equations in the 
presence of a constant active field background. One then applies Starobinsky's
formalism. This is also the same as for passive fields. The key difference is
that imposing a constant active field background does not induce a mass but
rather a field strength --- for nonlinear sigma models --- or a modification 
in the background geometry --- for differentiated actives coupled to gravity.

Another important point is that none of these theories is renormalizable.
However, all theories can be renormalized by subtracting BPHZ (Bogoliubov
and Parasiuk \cite{Bogoliubov:1957gp}, Hepp \cite{Hepp:1966eg} and
Zimmermann \cite{Zimmermann:1968mu,Zimmermann:1969jj}) counterterms,
order-by-order in the loop expansion. When this is done, the leading secular
growth factors are uniquely determined by low energy effective field theory
in the sense of Donoghue \cite{Donoghue:1993eb,Donoghue:1994dn,
Donoghue:2017ovt}.

\subsection{Nonlinear Sigma Models}

Quantum gravity is {\it hard}. The weeks it requires to perform a 2-loop
scalar computation on de Sitter background \cite{Onemli:2002hr,Onemli:2004mb,
Brunier:2004sb} become months to perform a 1-loop graviton computation 
\cite{Tsamis:1996qk,Park:2011ww,Miao:2024atw}. The fundamental $h \partial 
h \partial h$ interaction of quantum gravity also occurs in nonlinear sigma 
models such as,
\begin{eqnarray}
\mathcal{L}_1 &\!\!\! = \!\!\!& -\tfrac12 (1 + \tfrac{\lambda}{2} \phi)^2 
\partial_{\mu} \phi \partial_{\nu} \phi g^{\mu\nu} \sqrt{-g} \; , \qquad 
\label{1field} \\
\mathcal{L}_2 &\!\!\! = \!\!\!& -\tfrac12 \partial_{\mu} A \partial_{\nu} A
g^{\mu\nu} \sqrt{-g} - \tfrac12 (1 + \tfrac{\lambda}{2} A)^2 \partial_{\mu}
B \partial_{\nu} B g^{\mu\nu} \sqrt{-g} \; . \qquad \label{2field}
\end{eqnarray}
For this reason nonlinear sigma models have long been employed as a simple
venue to sort out the complexities of derivative interactions, without the
tensor indices, the gauge issue and the vast proliferation of interactions
which make quantum gravity so difficult \cite{Tsamis:2005hd,Kitamoto:2010et,
Kitamoto:2011yx,Kitamoto:2018dek,Miao:2021gic}.

The key to integrating out differentiated fields in (\ref{1field}-\ref{2field})
is the observation that working in a constant $\phi$ or $A$ background just 
changes the field strengths of the $\phi$ and $B$ propagators from that of
the massless, minimally coupled scalar $i\Delta(x;x')$,
\begin{equation}
\langle \Omega \vert T[ \phi(x) \phi(x') ] \vert \Omega \rangle_{\phi_0} =
\tfrac{i \Delta(x;x')}{(1 + \frac{\lambda}{2} \phi_0)^2} \quad , \quad
\langle \Omega \vert T[ B(x) B(x') ] \vert \Omega \rangle_{A_0} =
\tfrac{i \Delta(x;x')}{(1 + \frac{\lambda}{2} A_0)^2} \; . \label{phiBprops}
\end{equation}
Integrating out the $\partial B \partial B$ term (for constant $A_0$) from the 
$A$ field equation (and taking $D=4$) gives,
\begin{eqnarray}
\lefteqn{\partial_{\mu} [\sqrt{-g} \, g^{\mu\nu} \partial_{\nu} A] -
\tfrac{\lambda}{2} ( 1 + \tfrac{\lambda}{2} A) \partial_{\mu} B \partial_{\nu}
B g^{\mu\nu} \sqrt{-g} } \nonumber \\
& & \hspace{3.3cm} \longrightarrow \partial_{\mu} [\sqrt{-g} \, g^{\mu\nu} 
\partial_{\nu} A] - \tfrac{\lambda}{2} (1 + \tfrac{\lambda}{2} A_0) \times
-\tfrac{ \frac{3 H^4}{8 \pi^2} \sqrt{-g}}{(1 + \frac{\lambda}{2} A_0)^2} \; .
\qquad \label{Apotential}
\end{eqnarray}
Applying Starobinsky's formalism to this scalar potential model gives a
Langevin equation for the stochastic field $\mathcal{A}(t,\vec{x})$,
\begin{equation}
3 H (\dot{\mathcal{A}} - \dot{\mathcal{A}}_0) = \tfrac{3 \lambda H^4}{16 \pi^2}
\tfrac1{1 + \frac{\lambda}{2} \mathcal{A}} \; . \label{ALangevin}
\end{equation}

The effective potential for $A$ induced in (\ref{Apotential}) is 
$V(A) = -\frac{3 H^4}{8 \pi^2} \ln\vert 1 + \frac{\lambda}{2} A\vert$. This
is unbounded below so no static limit is approached. However, the time evolution
of the background is sedate. It consists of a ``classical'' contribution,
obtained by ignoring the jitter $\mathcal{A}_0$, plus a series in powers of
$\mathcal{A}_0$, 
\begin{eqnarray}
\lefteqn{ \mathcal{A}(t,\vec{x}) = -\tfrac{2}{\lambda} + \tfrac{2}{\lambda} 
\Bigl[1 + \tfrac{\lambda^2 H^2 \ln[a(t)]}{16 \pi^2}\Bigr]^{\frac12} } \nonumber \\
& & \hspace{1.5cm} + \mathcal{A}_0(t,\vec{x}) - \tfrac{\lambda^2 H^3}{32 \pi^2}
\!\! \int_{0}^{t} \!\!\!\! dt' \mathcal{A}_0(t',\vec{x}) + \tfrac{\lambda^3
H^3}{64 \pi^2} \!\! \int_{0}^{t} \!\!\!\! dt' \mathcal{A}_0^2(t',\vec{x})
+ O(\lambda^4) \; . \qquad \label{calAexp} 
\end{eqnarray}
Correlators of $\mathcal{A}$ agree, at leading logarithm order, with 
dimensionally regulated and fully BPHZ renormalized 1-loop and 2-loop 
computations of those of $A$, for example \cite{Miao:2021gic,Woodard:2023rqo},
\begin{equation}
\langle \Omega \vert \mathcal{A}(t,\vec{x}) \vert \Omega \rangle = 
\tfrac{\lambda H^2 \ln[a(t)]}{16 \pi^2} + \tfrac{\lambda^3 H^4 \ln^2[a(t)]}{2^{10} 
\pi^4} + O(\lambda^5) \; . \label{VEVA}
\end{equation}
Note that the basic time dependence of (\ref{VEVA}) derives from the classical
part on the first line of (\ref{calAexp}), and is understandable in terms of
the field rolling down its potential. The stochastic jitter accelerates the roll,
essentially because it is easier to fluctuate down a potential than up.

The stochastic formulation of the single field model (\ref{1field}) is crucial
for gravity because, in both cases, it is the {\it same} field whose derivatives
are integrated out and which experiences stochastic fluctuations. It is 
straightforward to integrate out differentiated scalars from the $\phi$ field 
equation,
\begin{eqnarray}
\lefteqn{ (1 + \tfrac{\lambda}{2} \phi) \partial_{\mu} \Bigl[ (1 + 
\tfrac{\lambda}{2} \phi) \sqrt{-g} \, g^{\mu\nu} \partial_{\nu} \phi\Bigr] }
\nonumber \\
& & \hspace{4cm} \longrightarrow (1 + \tfrac{\lambda}{2} \phi_0) \partial_{\mu}
\Bigl[ \tfrac{\lambda}{4} \sqrt{-g} \, g^{\mu\nu} \partial_{\nu} \langle \Omega
\vert \phi^2 \vert \Omega \rangle_{\phi_0} \Bigr] \; , \qquad \\
& & \hspace{4cm} \longrightarrow (1 + \tfrac{\lambda}{2} \phi_0) \times -
\tfrac{3 \lambda H^4}{16 \pi^2} \tfrac{\sqrt{-g}}{(1 + \frac{\lambda}{2} \phi_0)^2}
\; . \qquad \label{phipotential}
\end{eqnarray}
This corresponds to another logarithmic potential $V(\phi) = \frac{3 H^4}{8 \pi^2} 
\ln \vert 1 + \frac{\lambda}{2} \phi\vert$. The key point is that we add it to
the stochastic simplification of the same derivative terms from which it emerged,
\begin{equation}
3 H (1 + \tfrac{\lambda}{2} \varphi)^2 [\dot{\varphi} - \dot{\varphi}_0] =
-\tfrac{3 \lambda H^4}{16 \pi^2} \tfrac1{1 + \frac{\lambda}{2} \varphi} \; .
\label{varphiEQN}
\end{equation}
As with the double field model (\ref{calAexp}), the solution of this Langevin 
equation consists of a classical part plus a series in powers of $\varphi_0$,
\begin{eqnarray}
\lefteqn{ \varphi(t,\vec{x}) = -\tfrac{2}{\lambda} + \tfrac{2}{\lambda} 
\Bigl[1 - \tfrac{\lambda^2 H^2 \ln[a(t)]}{8 \pi^2}\Bigr]^{\frac14} } \nonumber \\
& & \hspace{1.5cm} + \varphi_0(t,\vec{x}) + \tfrac{3\lambda^2 H^3}{32 \pi^2}
\!\! \int_{0}^{t} \!\!\!\! dt' \varphi_0(t',\vec{x}) - \tfrac{3 \lambda^3
H^3}{32 \pi^2} \!\! \int_{0}^{t} \!\!\!\! dt' \varphi_0^2(t',\vec{x})
+ O(\lambda^4) . \qquad \label{varphiexp} 
\end{eqnarray}
Also like the double field model (\ref{VEVA}), correlators of $\varphi$ agree, at
leading logarithm order, with explicit 1-loop and 2-loop computations performed 
using dimensional regularization and BPHZ renormalization \cite{Miao:2021gic},
\begin{equation}
\langle \Omega \vert \varphi(t,\vec{x}) \vert \Omega \rangle = - 
\tfrac{\lambda H^2 \ln[a(t)]}{16 \pi^2} - \tfrac{15 \lambda^3 H^4 \ln^2[a(t)]}{2^{10} 
\pi^4} + O(\lambda^5) \; . \label{VEVphi}
\end{equation}

\subsection{Scalar Corrections to Gravity}

Most matter fields are not active, but the massless, minimally coupled scalar
is an exception,
\begin{equation}
\mathcal{L}_{\scriptscriptstyle \rm MMC} = -\tfrac12 \partial_{\mu} \phi
\partial_{\nu} \phi g^{\mu\nu} \sqrt{-g} \; . \label{MMC}
\end{equation}
To integrate differentiated scalar out of the gravitational field equation,
\begin{equation}
R_{\mu\nu} - \tfrac12 g_{\mu\nu} R + (\tfrac{D-2}{2}) \Lambda g_{\mu\nu} =
8\pi G \Bigl\{ \partial_{\mu} \phi \partial_{\nu} \phi - \tfrac12 g_{\mu\nu}
g^{\rho\sigma} \partial_{\rho} \phi \partial_{\sigma} \phi \Bigr\} \; ,
\label{MMCEinstein}
\end{equation}
we must define what it means to be in a ``constant graviton background''.
If the temporal coordinate is changed from co-moving time $t$ to conformal
time $\eta$, such that $d\eta = dt/a(t)$, the background geometry of
cosmology (\ref{geometry}) takes the form $a^2 \eta_{\mu\nu}$. We define
the graviton field $h_{\mu\nu}(x)$ by conformally transforming the full
metric,
\begin{equation}
g_{\mu\nu} \equiv a^2 \widetilde{g}_{\mu\nu} \equiv a^2 (\eta_{\mu\nu} +
\kappa h_{\mu\nu}) \qquad , \qquad \kappa^2 \equiv 16 \pi G \; . \label{hdef}
\end{equation}
The transverse-traceless and purely spatial components of $h_{\mu\nu}$ obey 
the same equation of motion as that of the massless, minimally coupled scalar 
\cite{Lifshitz:1945du}. Consequently, it is $h_{\mu\nu}$ which experiences 
inflationary particle production (\ref{occupation}), and $h_{\mu\nu}$ which
should be held constant when we integrate out differentiated scalars from
equation (\ref{MMCEinstein}). 

Constant $h_{\mu\nu}$ means that $\widetilde{g}_{\mu\nu}$ is also constant. 
It turns out that setting $g_{\mu\nu} = a^2 \widetilde{g}_{\mu\nu}$, with 
constant $\widetilde{g}_{\mu\nu}$ amounts to being in de Sitter with a 
different Hubble parameter \cite{Basu:2016iua,Basu:2016gyg},
\begin{equation}
\partial_{\rho} \widetilde{g}_{\mu\nu} = 0 \qquad \Longrightarrow \qquad
H^2 \longrightarrow -\widetilde{g}^{00} H^2 \equiv \widetilde{H}^2 \; .
\label{Htilde}
\end{equation}
This makes it simple to integrate out the scalars from equation
(\ref{MMCEinstein}),
\begin{eqnarray}
\lefteqn{ \tfrac{\kappa^2}{2} \Bigl\{ \partial_{\mu} \phi \partial_{\nu} 
\phi \!-\! \tfrac12 g_{\mu\nu} g^{\rho\sigma} \partial_{\rho} \phi 
\partial_{\sigma} \phi \Bigr\} \!\longrightarrow\! \tfrac{\kappa^2}{2} 
\Bigl( \delta^{\rho}_{~\mu} \delta^{\sigma}_{~\nu} \!-\! \tfrac12 g_{\mu\nu} 
g^{\rho\sigma} \Bigr) \!\times\! \langle \Omega \vert \partial_{\rho} \phi
\partial_{\sigma} \phi \vert \Omega \rangle_{h} \; , } \\
& & \hspace{3cm} \longrightarrow \tfrac{\kappa^2}{2} \Bigl( 
\delta^{\rho}_{~\mu} \delta^{\sigma}_{~\nu} - \tfrac12 g_{\mu\nu} 
g^{\rho\sigma} \Bigr) \times -\tfrac{3 \widetilde{H}^4}{32 \pi^2} \, 
g_{\rho\sigma} = \tfrac{3 \kappa^2 \widetilde{H}^4}{64 \pi^2} \, g_{\mu\nu} 
\; . \qquad \label{MMCstress}
\end{eqnarray}
Expression (\ref{MMCstress}) corresponds to a finite renormalization of the
cosmological constant \cite{Miao:2024nsz},
\begin{equation}
\delta \Lambda = -\tfrac{3 \kappa^2 \widetilde{H}^4}{64 \pi^2} \; .
\label{deltaLambda}
\end{equation}
The need for this renormalization to make the graviton self-energy
conserved had been discovered previously through explicit computations
\cite{Tsamis:2023fri,Miao:2024atw}.

\subsection{Pure Gravity}

There is no more postponing the complexities of gravity. The vast proliferation
of interactions is apparent from expressing the invariant Lagrangian in terms
of the graviton field $h_{\mu\nu}$ and the conformally transformed metric
$\widetilde{g}_{\mu\nu} \equiv \eta_{\mu\nu} + \kappa h_{\mu\nu}$
\cite{Tsamis:1992xa},
\begin{eqnarray}
\lefteqn{\mathcal{L}_{\rm inv} = a^{D-2} \sqrt{-\widetilde{g}} \, 
\widetilde{g}^{\alpha\beta} \widetilde{g}^{\rho\sigma} \widetilde{g}^{\mu\nu} 
\Bigl\{ \tfrac12 h_{\alpha\rho, \mu} h_{\nu\sigma, \beta} - \tfrac12 
h_{\alpha\beta, \rho} h_{\sigma\mu, \nu} } \nonumber \\
& & \hspace{.5cm} + \tfrac14 h_{\alpha\beta, \rho} h_{\mu\nu, \sigma} 
- \tfrac14 h_{\alpha\rho, \mu} h_{\beta\sigma, \nu} \Bigr\}  
+ (\tfrac{D-2}{2}) a^{D-1} H \sqrt{-\widetilde{g}} \, 
\widetilde{g}^{\rho\sigma} \widetilde{g}^{\mu\nu} h_{\rho\sigma, \mu} 
h_{\nu 0} \; . \qquad \label{invariant}
\end{eqnarray}
The simplest gauge is based on adding the gauge fixing term \cite{Tsamis:1992xa,
Woodard:2004ut},
\begin{equation}
\mathcal{L}_{\rm GF} =-\tfrac12 a^{D-2} \eta^{\mu\nu} F_{\mu} F_{\nu} \;\; ,
\;\; F_{\mu} = \eta^{\rho\sigma} [h_{\mu\rho ,\sigma} - \tfrac12 
h_{\rho\sigma , \mu} + (D\!-\!2) a H h_{\mu\rho} \delta^0_{~\sigma}] \; .
\label{gaugefixing}
\end{equation}
The associated ghost Lagrangian is,
\begin{equation}
\mathcal{L}_{\rm gh} = -a^{D-2} \eta^{\mu\nu} \overline{c}_{\mu} \delta
F_{\nu} \; , \label{ghost}
\end{equation}
where $\delta F_{\nu}$ represents the infinitesimal transformation $x^{\mu} 
\rightarrow x^{\mu} + \kappa \epsilon^{\mu}$ of $F_{\nu}$ with the transformation
parameter replaced by the ghost field.

The gauge (\ref{gaugefixing}-\ref{ghost}) results in ghost and graviton
propagators whose tensor structure consists of constants formed from 
$\eta_{\mu\nu}$ and $\delta^0_{~\mu}$,
\begin{eqnarray}
i [\mbox{}_{\mu\nu} \Delta_{\rho\sigma}](x;x') &\!\!\! = \!\!\!& \sum_{I=A,B,C}
[\mbox{}_{\mu\nu} T^{I}_{\rho\sigma}] \times i\Delta_I(x;x') \; , \qquad
\label{gravprop} \\
i [\mbox{}_{\mu} \Delta_{\rho}](x;x') &\!\!\! = \!\!\!& \overline{\eta}_{\mu\nu}
\times i\Delta_A(x;x') - \delta^0_{~\mu} \delta^0_{~\nu} \times i \Delta_B(x;x')
\; . \qquad \label{ghostprop}
\end{eqnarray}
Here and henceforth the purely spatial Minkowski metric is $\overline{\eta}_{\mu\nu} 
\equiv \eta_{\mu\nu} + \delta^0_{~\mu} \delta^0_{\nu}$, and the other tensor 
factors are,
\begin{eqnarray}
[\mbox{}_{\mu\nu} T^A_{\rho\sigma}] &\!\!\! = \!\!\!& 2\overline{\eta}_{\mu (\rho} 
\overline{\eta}_{\sigma) \nu} - \tfrac{2}{D-3} \overline{\eta}_{\mu\nu}
\overline{\eta}_{\rho\sigma} \qquad , \qquad [\mbox{}_{\mu\nu} T^B_{\rho\sigma}]
= -4 \delta^0_{~(\mu} \overline{\eta}_{\nu) (\rho} \delta^0_{~\sigma)} \; , 
\qquad \label{TATB} \\
{[} \mbox{}_{\mu\nu} T^C_{\rho\sigma} {]} 
&\!\!\! = \!\!\!& \tfrac{2}{(D-2) (D-3)} 
[(D\!-\!3) \delta^0_{~\mu} \delta^0_{~\nu} + \overline{\eta}_{\mu\nu}] 
[(D\!-\!3) \delta^0_{~\rho} \delta^0_{~\sigma} + \overline{\eta}_{\rho\sigma}]
\; . \qquad \label{TC}
\end{eqnarray}
Parenthesized indices are symmetrized. The three propagators $i\Delta_{I}(x;x')$ 
are those of a minimally coupled scalar with masses,
\begin{equation}
m_A^2 = 0 \qquad , \qquad m_B^2 = (D\!-\!2) H^2 \qquad , \qquad m_C^2 = 2
(D\!-\!3) H^2 \; . \label{scalarmasses}
\end{equation}

Because graviton propagators possess indices, working in a constant 
$\widetilde{g}_{\mu\nu}$ background requires more than just replacing all the
factors of $H^2$ by $\widetilde{H}^2$, as per equation (\ref{Htilde}), in the 
scalar propagators $i\Delta_I(x;x')$. One must also modify the gauge fixing
function to \cite{Miao:2024shs},
\begin{equation}
\widetilde{\mathcal{L}}_{\rm GF} =-\tfrac12 a^{D-2} \sqrt{-\widetilde{g}} \,
\widetilde{g}^{\mu\nu} \widetilde{F}_{\mu} \widetilde{F}_{\nu} \;\; ,
\;\; \widetilde{F}_{\mu} = \widetilde{g}^{\rho\sigma} [h_{\mu\rho ,\sigma} \!-\! 
\tfrac12 h_{\rho\sigma , \mu} \!+\! (D\!-\!2) a H h_{\mu\rho} \delta^0_{~\sigma}] 
\; . \label{newgauge}
\end{equation}
The ghost Lagrangian (\ref{ghost}) suffers a similar change \cite{Miao:2024shs},
\begin{equation}
\widetilde{\mathcal{L}}_{\rm gh} = -a^{D-2} \sqrt{-\widetilde{g}} \,
\widetilde{g}^{\mu\nu} \overline{c}_{\mu} \delta \widetilde{F}_{\nu} \; . 
\label{newghost}
\end{equation}

The tensor structures of the ghost and graviton propagators 
(\ref{gravprop}-\ref{TC}) are constructed from the Minkowski metric 
$\eta_{\mu\nu}$ and the timelike 4-vector $\delta^0_{~\mu}$. Generalizing
to a constant $\widetilde{g}_{\mu\nu}$ background amounts to stating what
becomes of these two tensors. The most straightforward is $\eta_{\mu\nu}$:
it generalizes to $\widetilde{g}_{\mu\nu}$. Explaining what happens to 
$\delta^0_{~\mu}$ is facilitated by the ADM decomposition 
\cite{Arnowitt:1962hi},
\begin{equation}
\widetilde{g}_{\mu\nu} dx^{\mu} dx^{\nu} = -N^2 d\eta^2 + \gamma_{ij}
(dx^i - N^i d\eta) (dx^j - N^j d\eta) \; . \label{ADM}
\end{equation}
This suggests the generalization \cite{Miao:2024shs},
\begin{eqnarray}
\delta^0_{~\mu} &\!\!\! \longrightarrow \!\!\!& u_{\mu} \equiv -N 
\delta^{0}_{~\mu} \; , \qquad \label{timelike} \\
\overline{\eta}_{\mu\nu} &\!\!\! \longrightarrow \!\!\!& 
\overline{\gamma}_{\mu\nu} \equiv \widetilde{g}_{\mu\nu} + u_{\mu} u_{\nu} 
\; . \qquad \label{etabar}
\end{eqnarray}
So the graviton and ghost propagators in a constant $\widetilde{g}_{\mu\nu}$
background are,
\begin{eqnarray}
i [\mbox{}_{\mu\nu} \widetilde{\Delta}_{\rho\sigma}](x;x') &\!\!\! = \!\!\!& 
\sum_{I=A,B,C} [\mbox{}_{\mu\nu} \widetilde{T}^{I}_{\rho\sigma}] \times i
\widetilde{\Delta}_I(x;x') \; , \qquad \label{gravtilde} \\
i [\mbox{}_{\mu} \widetilde{\Delta}_{\rho}](x;x') &\!\!\! = \!\!\!& 
\overline{\gamma}_{\mu\nu} \times i\widetilde{\Delta}_A(x;x') - u_{\mu} u_{\nu} 
\times i \widetilde{\Delta}_B(x;x') \; . \qquad \label{ghosttilde}
\end{eqnarray}
The scalar propagators $i \widetilde{\Delta}_{I}(x;x')$ are obtained by
making the replacement (\ref{Htilde}) in $i \Delta_I(x;x')$, and the tensor
factors $[\mbox{}_{\mu\nu} \widetilde{T}^I_{\rho\sigma}]$ are obtained by
making the replacements (\ref{timelike}-\ref{etabar}) in (\ref{TATB}-\ref{TC}).

The sum of the invariant Lagrangian (\ref{invariant}) and the new gauge fixing
term (\ref{newgauge}) can be grouped into six terms \cite{Miao:2024shs},
\begin{eqnarray}
\mathcal{L}_{\scriptscriptstyle 1+2+3} &\!\!\! = \!\!\!& a^{D-2} 
\sqrt{-\widetilde{g}} \, \widetilde{g}^{\alpha\beta} [-\tfrac14 
\widetilde{g}^{\gamma\rho} \widetilde{g}^{\delta\sigma} + \tfrac18
\widetilde{g}^{\gamma\delta} \widetilde{g}^{\rho\sigma}] 
h_{\gamma\delta , \alpha} h_{\rho\sigma , \beta} \nonumber \\
& & \hspace{3.5cm} + (\tfrac{D-2}{2}) a^D \sqrt{-\widetilde{g}} \, 
\widetilde{g}^{\gamma\alpha} \widetilde{g}^{\beta\rho} 
\widetilde{g}^{\sigma\delta} h_{\alpha\beta} h_{\rho\sigma} 
\widetilde{H}^2 u_{\gamma} u_{\delta} \; , \qquad \label{L123} \\
\mathcal{L}_{\scriptscriptstyle 4+5} &\!\!\! = \!\!\!& 
\sqrt{-\widetilde{g}} \, \widetilde{g}^{\alpha\beta} 
\widetilde{g}^{\gamma\delta} \widetilde{g}^{\rho\sigma} \partial_{\beta}
\Bigl[ -\tfrac12 \partial_{\sigma} (a^{D-2} h_{\gamma\rho} 
h_{\delta\alpha}) + a^{D-2} h_{\gamma\rho} h_{\delta \alpha , \sigma}
\Bigr] \; , \qquad \label{L45} \\
\mathcal{L}_{\scriptscriptstyle 6} &\!\!\! = \!\!\!& (\tfrac{D-2}2)
\kappa H a^{D-1} \sqrt{-\widetilde{g}} \, \widetilde{g}^{\alpha\beta}
\widetilde{g}^{\gamma\delta} \widetilde{g}^{\rho\sigma} h_{\rho\sigma ,
\gamma} h_{\delta\alpha} h_{\beta 0} \; . \qquad \label{L6}
\end{eqnarray}
Completing the stochastic simplification of pure gravity entails three
steps:
\begin{enumerate*}
\item{Find the contribution to the equation of motion from (\ref{newghost})
and (\ref{L123}-\ref{L6});}
\item{Integrate out the ghost fields and the differentiated graviton fields 
from each contribution; and}
\item{Derive the Langevin kinetic operator from each contribution.}
\end{enumerate*}
The final step involves distinguishing between the spatial components of
the graviton field, which are active, and the other components, which 
are passive, constrained fields whose dynamics is driven by the active
components \cite{Woodard:2025smz},
\begin{equation}
\kappa h_{\mu\nu} = A_{\mu\nu} + 2 u_{(\mu} B_{\nu)} + [u_{\mu} u_{\nu}
+ \overline{\gamma}_{\mu\nu}] C \qquad , \qquad u^{\rho} A_{\rho\sigma}
= 0 = u^{\rho} B_{\rho} \; . \label{distinguish}
\end{equation}
Note that the active components $A_{\mu\nu}$ have a stochastic jitter
$a_{\mu\nu}$ whereas the passive, constrained fields $B_{\mu}$ and $C$ do 
not.

The first two steps have been implemented \cite{Miao:2024shs}, and the 
third is in progress \cite{Woodard:2025smz}. The complete result for
(\ref{L123}) is,
\begin{eqnarray}
\lefteqn{ \tfrac{\kappa a^{-4}}{\sqrt{-\widetilde{g}}} \!\times\! 
\tfrac{\delta S_{1+2+3}}{\delta h_{\mu\nu} } \longrightarrow 
\tfrac{\kappa^2 \widetilde{H}^4}{8 \pi^2} [13 \widetilde{g}^{\mu\nu}
\!\!\!+\! 6 u^{\mu} u^{\nu}] \!-\! [\widetilde{g}^{\mu \rho} 
\widetilde{g}^{\nu \sigma} \!\!\!-\! \tfrac12 \widetilde{g}^{\mu\nu} 
\widetilde{g}^{\rho\sigma} ] a^{-1} \! \widetilde{H} u^{\alpha} 
\partial_{\alpha} (A_{\rho\sigma} \!\!-\! a_{\rho\sigma}) } \nonumber \\
& & \hspace{0cm} + u^{(\mu} \widetilde{g}^{\nu) \rho} \widetilde{D}_{B} 
B_{\rho} + u^{\mu} u^{\nu} \widetilde{D}_C C + \widetilde{H}^2 \Bigl\{ 
\tfrac12 \widetilde{g}^{\mu\nu} (B^2 \!-\! C^2) \nonumber \\
& & \hspace{2.5cm}  \!+\! 2 u^{(\mu} A^{\nu) \rho} B_{\rho} \!-\! 
B^{\mu} B^{\nu} \!-\! 2 u^{(\mu} B^{\nu)} C \!+\! u^{\mu} u^{\nu} 
(2 B^2 \!-\! 3 C^2) \Bigr\} \; . \qquad \label{reduced123}
\end{eqnarray}
Although the two operators agree for $D=4$, it is useful to retain 
dimensional regularization in order to see the difference between
$\widetilde{D}_{B}$ and $\widetilde{D}_C$,
\begin{eqnarray}
\widetilde{D}_B \!\!\!& = \!\!\!& \partial_{\alpha} [a^{D-2} 
\sqrt{-\widetilde{g}} \, \widetilde{g}^{\alpha\beta} \partial_{\beta}]
- (D\!-\!2) \widetilde{H}^2 a^D \sqrt{-\widetilde{g}} \; , \qquad
\label{DB} \\
\widetilde{D}_C \!\!\!& = \!\!\!& \partial_{\alpha} [a^{D-2} 
\sqrt{-\widetilde{g}} \, \widetilde{g}^{\alpha\beta} \partial_{\beta}]
- 2(D\!-\!3) \widetilde{H}^2 a^D \sqrt{-\widetilde{g}} \; . \qquad
\label{DC} 
\end{eqnarray}
In $D=4$ both operators degenerate to the conformal d'Alembertian.

Whereas nonlinear sigma models and scalar corrections to gravity had
seen extensive dimensionally regulated and fully renormalized computations 
prior to their stochastic realizations, the stochastic formulation of pure 
gravity is far ahead of the explicit calculations with which it might be
compared. The 1PI (one-particle-irreducible) 1-graviton function has been 
computed in the original gauge (\ref{gaugefixing}) \cite{Tsamis:2005je}, 
but it must be re-done in the new gauge (\ref{newgauge}) to check the 
stochastic equation. Quantum gravity presumably requires the same sort of 
finite renormalization of the cosmological constant that was found for 
scalar corrections to gravity \cite{Tsamis:2023fri}. Also in the old gauge 
(\ref{gaugefixing}), the 1PI 2-point function (the ``graviton self-energy'') 
was computed, away from coincidence and without using dimensional 
regularization or renormalizing \cite{Tsamis:1996qk}. This needs to re-done 
in the new gauge (\ref{newgauge}) using dimensional regularization and BPHZ 
renormalization. A technique was devised for extending the old calculation 
to a fully renormalized result \cite{Tan:2021ibs}. This extension was then 
employed to compute 1-loop corrections to the graviton mode function 
\cite{Tan:2021lza} and to the gravitational response to a point mass 
\cite{Tan:2022xpn}. Both of these things need to re-done with a exact
computation in the new gauge.

\section{The Other Source of Large Logarithms}

The first large logarithm from a loop of gravitons was discovered 20 years
ago \cite{Miao:2005am,Miao:2006gj}. It was soon shown not to follow from a
naive realization of the stochastic formalism \cite{Miao:2008sp}. In the 
meantime a number of similar effects were found from graviton loop 
corrections to matter \cite{Kahya:2007bc,Leonard:2013xsa,Glavan:2013jca,
Wang:2014tza,Glavan:2020gal,Glavan:2020ccz,Glavan:2021adm} and in graviton
corrections to gravity \cite{Tan:2021ibs,Tan:2021lza,Tan:2022xpn}. The same
kinds of derivative interactions in matter corrections to gravity also 
induce large logarithms \cite{Wang:2015eaa,Miao:2024atw,Foraci:2024vng,
Foraci:2024cwi}. Explaining these effects, and devising techniques to
resum them, has been a long and confusing struggle for two reasons:
\begin{enumerate*}
\item{The correct stochastic formulation involves integrating differentiated
active fields out, which entails new sorts of effective potentials based on
background-dependent field strengths and Hubble parameters; and}
\item{There is another way large logarithms can be induced which cannot be
captured stochastically.}
\end{enumerate*}
Section 5 concerned the first problem, this section concerns the second.

The second way of inducing large logarithms is the incomplete cancellation 
between primitive divergences and counterterms. When using dimensional
regularization it turns out that the factors of $a^{D-2}$ from vertices
are canceled by inverse factors from propagators. For example, the 
$i\Delta_I(x;x')$ propagators which appear in the graviton and ghost 
propagators (\ref{gravprop}-\ref{ghostprop}) have the same leading 
singularity,
\begin{equation}
i \Delta_I(x;x') = \tfrac1{4 \pi^{\frac{D}2}} \tfrac1{[a a' (x - x')^2]^{
\frac{D}2 - 1}} + \dots \; , \label{leading}
\end{equation}
where $(x - x')^2 \equiv (x-x')^{\mu} (x-x')^{\nu} \eta_{\mu\nu}$ is the
conformal coordinate interval. Higher terms in the propagator can give
rise to additional, integer powers of $(a a')$, but the $D$-dependence
will always be $(a a')^{-\frac{D}2}$. On the other hand, counterterms
inherit a factor of $a^D$ from the $\sqrt{-g}$ measure. It follows that
there is an incomplete cancellation between primitive divergences and
counterterms,
\begin{equation}
\tfrac{(2 H)^{D-4}}{D-4} - \tfrac{(a \mu)^{D-4}}{D-4} = -
\ln(\tfrac{a \mu}{2 H}) + O(D \!-\! 4) \; . \label{renorm}
\end{equation}

Nonlinear sigma models exhibit both stochastic logarithms such as
(\ref{VEVA}) and (\ref{VEVphi}), and also renormalization-induced 
logarithms (\ref{leading}). The large logarithms of most graviton
loop corrections to matter fields \cite{Kahya:2007bc,Leonard:2013xsa,
Glavan:2013jca,Wang:2014tza,Glavan:2021adm}, and all known matter
loop corrections to gravity \cite{Park:2011ww,Wang:2015eaa,Miao:2024atw,
Foraci:2024vng,Foraci:2024cwi}, are induced by renormalization. The case 
of pure gravity \cite{Tan:2021ibs,Tan:2021lza,Tan:2022xpn} has yet to 
be decided.

Renormalization-induced logarithms occur even for passive fields which
do not experience inflationary particle production (\ref{occupation}),
so they cannot be explained stochastically. On the other hand, the 
close association in expression (\ref{renorm}) between the dimensional
regularization scale $\mu$ and the scale factor $a(t)$ makes these
logarithms amenable to a modified renormalization group treatment. I
will explain the technique in the context of matter loop corrections
to gravity \cite{Park:2011ww,Wang:2015eaa,Miao:2024atw,Foraci:2024vng,
Foraci:2024cwi} for which a unified treatment is possible.

What has been done is to compute matter loop contributions to the 
graviton self-energy $-i [\mbox{}^{\mu\nu} \Sigma^{\rho\sigma}](x;x')$
on de Sitter background \cite{Park:2011ww,Wang:2015eaa,Miao:2024atw,
Foraci:2024vng,Foraci:2024cwi}. This can be used to quantum-correct 
the linearized Einstein equation,
\begin{equation}
\mathcal{L}^{\mu\nu\rho\sigma} \kappa h_{\rho\sigma} -
\int \!\! d^4x' \, [\mbox{}^{\mu\nu} \Sigma^{\rho\sigma}](x;x') \kappa 
h_{\rho\sigma}(x') = \tfrac{\kappa^2}{2} T^{\mu\nu}(x) \; . \label{QEin}
\end{equation}
Here $\mathcal{L}^{\mu\nu\sigma\rho}$ is the Lichnerowicz operator on 
de Sitter background and $T^{\mu\nu}$ is the ``graviton stress tensor,''
defined as minus the variation of the matter action with respect to
$h_{\mu\nu}$. With $T^{\mu\nu} = 0$, one is solving for 1-loop 
corrections to gravitational radiation. There are no corrections in 
flat space background, but a loop of massless, minimally coupled 
scalars gives rise to a secular enhancement of the Weyl tensor 
\cite{Miao:2024atw},
\begin{equation}
C_{0i0j}(t,\vec{x}) = C^{\rm tree}_{0i0j}(t,\vec{x}) \Bigl\{1 -
\tfrac{3 \kappa^2 H^2}{160 \pi^2} \ln[a(t)] + \dots \Bigr\} \; .
\label{MMCWeyl} 
\end{equation}
Setting $T^{\mu\nu}(t,\vec{x}) = -M \delta^{\mu}_{~0} \delta^{\nu}_{~0}
a(t) \delta^3(\vec{x})$ gives the response to a point mass, which can
be parameterized in terms of two scalar potentials,
\begin{equation}
ds^2 = -[1 - 2 \Psi(t,r)] dt^2 + a^2(t) [1 - 2 \Phi(t,r)] d\vec{x} \cdot
d\vec{x} \; . \label{scalarPs}
\end{equation}
For a loop of massless, minimally coupled scalars one finds 
\cite{Park:2015kua,Miao:2024atw},
\begin{eqnarray}
\Psi(t,r) &\!\!\! = \!\!\!& \tfrac{GM}{a(t) r} \Bigl\{1 + \tfrac{\kappa^2}{
320 \pi^2 a^2(t) r^2} - \tfrac{3 \kappa^2 H^2}{160 \pi^2} \ln[a(t)] + \dots
\Bigr\} \; , \qquad \label{MMCNewton} \\
\Psi(t,r) + \Phi(t,r) &\!\!\! = \!\!\!& \tfrac{GM}{a(t) r} \Bigl\{0 + 
\tfrac{\kappa^2}{240 \pi^2 a^2(t) r^2} + \tfrac{3 \kappa^2 H^2}{160 \pi^2}
+ \dots \Bigr\} \; . \qquad \label{MMCSlip}
\end{eqnarray}
The fractional $\kappa^2/a^2r^2$ correction in (\ref{MMCNewton}) is
the de Sitter descendant of an old flat space effect 
\cite{Radkowski:1970}, whereas the $\kappa^2 H^2$ corrections are new.
Corrections to the Weyl tensor and to the two potentials have also been 
computed for conformally invariant matter theories. For a massless,
conformally coupled scalar the results are \cite{Foraci:2024cwi},
\begin{eqnarray}
C_{0i0j}(t,\vec{x}) &\!\!\! = \!\!\!& C^{\rm tree}_{0i0j}(t,\vec{x}) 
\Bigl\{1 + \tfrac{\kappa^2 H^2}{480 \pi^2} \ln[a(t)] + \dots \Bigr\} \; , 
\qquad \label{MCCWeyl} \\
\Psi(t,r) &\!\!\! = \!\!\!& \tfrac{GM}{a(t) r} \Bigl\{1 + \tfrac{\kappa^2}{
720 \pi^2 a^2(t) r^2} + \tfrac{\kappa^2 H^2}{480 \pi^2} \ln[a(t)] + \dots
\Bigr\} \; , \qquad \label{MCCNewton} \\
\Psi(t,r) + \Phi(t,r) &\!\!\! = \!\!\!& \tfrac{GM}{a(t) r} \Bigl\{0 + 
\tfrac{\kappa^2}{1440 \pi^2 a^2(t) r^2} - \tfrac{\kappa^2 H^2}{480 \pi^2}
+ \dots \Bigr\} \; . \qquad \label{MCCSlip}
\end{eqnarray}
As with its minimally coupled cousin, the $\kappa^2/a^2 r^2$ 
corrections to (\ref{MCCNewton}-\ref{MCCSlip}) are well known from
flat space background \cite{Capper:1973bk}. All (new and old) of the 
1-loop corrections from massless, Dirac fermions are a factor of $6$ 
times those in (\ref{MCCWeyl}-\ref{MCCSlip}) \cite{Capper:1973mv,
Foraci:2024cwi}. For a loop of photons the factor is $12$ 
\cite{Duff:2000mt,Capper:1974ed,Wang:2015eaa,Foraci:2024vng}.

The Lagrangian of general relativity with a cosmological constant
$\Lambda$ is,
\begin{equation}
\mathcal{L}_{\rm GR} = \tfrac{[R - (D-2) \Lambda] \sqrt{-g}}{16 \pi G} 
\; . \label{GR}
\end{equation}
On the other hand, all single matter loop corrections to gravity require
the same two counterterms \cite{tHooft:1974toh},
\begin{equation}
\Delta \mathcal{L} = c_1 R^2 \sqrt{-g} + c_2 C^{\alpha\beta\gamma\delta}
C_{\alpha\beta\gamma\delta} \sqrt{-g} \; . \label{cterms}
\end{equation}
The first problem in applying the renormalization group to general
relativity (\ref{GR}) is that the counterterms (\ref{cterms}) do not 
seem to represent renormalizations of the classical action. However,
one can easily decompose the Eddington ($R^2$) counterterm into a
higher derivative part and two lower derivative parts which can be 
viewed as part of the classical action \cite{Miao:2024nsz},
\begin{equation}
R^2 = [R \!-\! D \Lambda \!+\! D \Lambda]^2 = (R \!-\! D \Lambda)^2 +
2 D \Lambda [R \!-\! (D\!-\!2) \Lambda] + (D\!-\!4) D \Lambda^2 \; .
\label{Eddington}
\end{equation}
The final term in (\ref{Eddington}) could be regarded as renormalizing 
the cosmological constant, but the factor of $(D-4)$ makes it irrelevant.
In contrast, the middle term of (\ref{Eddington}) could be regarded as 
a renormalization of the graviton field strength.

Making a similar decomposition of the Weyl ($C^{\alpha\beta\gamma\delta}
C_{\alpha\beta\gamma\delta}$) counterterm is more challenging. We first
express it in terms of the Gauss-Bonnet scalar $G$,
\begin{eqnarray}
C^{\mu\nu\rho\sigma} C_{\mu\nu\rho\sigma} &\!\!\! = \!\!\!& 
R^{\mu\nu\rho\sigma} R_{\mu\nu\rho\sigma} -\tfrac{4}{D-2} R^{\mu\nu} 
R_{\mu\nu} + \tfrac{2 R^2}{(D-1) (D-2)} \; , \qquad \\
&\!\!\! = \!\!\!& G + 4 (\tfrac{D-3}{D-2}) R^{\mu\nu} R_{\mu\nu} -
\tfrac{D (D-3)}{(D-1) (D-2)} R^2 \; . \qquad \label{Weyl}
\end{eqnarray}
The Gauss-Bonnet scalar is,
\begin{equation}
G \equiv R^{\mu\nu\rho\sigma} R_{\mu\nu\rho\sigma} - 4 R^{\mu\nu} R_{\mu\nu}
+ R^2 \; . \label{GaussBonnet}
\end{equation}
It is significant by virtue of being a total derivative in $D=4$ dimensions,
which means it cannot affect 1-loop divergences. We now decompose the square
of the Ricci tensor similar to (\ref{Eddington}) into a higher derivative
term and two lower derivatives,
\begin{equation}
R^{\mu\nu} R_{\mu\nu} = (R^{\mu\nu} \!-\! \Lambda g^{\mu\nu}) (R_{\mu\nu} 
\!-\! \Lambda g_{\mu\nu}) + 2 \Lambda [R \!-\! (D\!-\!2) \Lambda] +
(D \!-\! 4) \Lambda^2 \; . \label{Ricci}
\end{equation}
Substituting (\ref{Eddington}) and (\ref{Ricci}) in equation (\ref{Weyl})
gives,
\begin{eqnarray}
\lefteqn{C^{\mu\nu\rho\sigma} C_{\mu\nu\rho\sigma} = 4 (\tfrac{D-3}{D-2})
(R^{\mu\nu} \!-\! \Lambda g^{\mu\nu}) (R_{\mu\nu} \!-\! \Lambda g_{\mu\nu})
- \tfrac{D (D-3)}{(D-1) (D-2)} (R \!-\! D \Lambda)^2 } \nonumber \\
& & \hspace{2cm} -\tfrac{2 (D-2) (D-3)}{D-1} \Lambda [R \!-\! (D\!-\!2) 
\Lambda)] - \tfrac{(D-2) (D-3) (D-4)}{D-1} \Lambda^2 + G \; . \qquad 
\label{Weylsimp} 
\end{eqnarray}
 
Expressions (\ref{Eddington}) and (\ref{Weylsimp}) imply that the 
counterterms (\ref{cterms}) can be viewed as a graviton field strength
renormalization, which gives rise to a gamma function,
\begin{equation}
\delta Z = 2 \Bigl[D (D\!-\!1) c_1 - (D\!-\!2) (D\!-\!3) c_2\Bigr] 
\kappa^2 H^2 \qquad \Longrightarrow \qquad \gamma = \tfrac{\partial
\ln(1 + \delta Z)}{\partial \ln(\mu^2)} \; . \label{gamma}
\end{equation}
The linearized Weyl tensor and the Newtonian potential can be viewed as 
2-point Green's functions, for which the Callan-Symanzik equation implies,
\begin{equation}
\Bigl[ \tfrac{\partial}{\partial \ln(\mu)} + 2 \gamma \Bigr] G^{(2)} = 0 \; .
\label{CSeqn}
\end{equation}
Relation (\ref{renorm}) shows that factors of $\ln(\mu)$ always come in 
the form $\ln(\mu a)$, so it should be valid to replace the derivative
with respect to $\ln(\mu)$ in expression (\ref{CSeqn}) by the derivative
with respect to $\ln(a)$. At this point one sees that the it is possible 
to predict the large temporal logarithms from renormalization. 

The results are impressive. For a loop of massless, minimally coupled
scalars the constants $c_1$ and $c_2$ in (\ref{cterms}) are
\cite{Park:2011ww,Miao:2024atw},
\begin{eqnarray}
c_1 &\!\!\! = \!\!\!& \tfrac{\mu^{D-4} \Gamma(\frac{D}2)}{2^8 \pi^{\frac{D}2}}
\tfrac{(D-2)}{(D-1)^2 (D-3) (D-4) } \; , \qquad \label{MMCc1} \\
c_2 &\!\!\! = \!\!\!& \tfrac{\mu^{D-4} \Gamma(\frac{D}2)}{2^8 \pi^{\frac{D}2}}
\tfrac{2}{(D+1) (D-1)^2 (D-3)^2 (D-4) } \; . \qquad \label{MMCc2}
\end{eqnarray}
Substituting these results in expression (\ref{gamma}) give the gamma function,
\begin{equation}
\gamma_{\rm \scriptscriptstyle MMCS} = \tfrac{3 \kappa^2 H^2}{320 \pi^2} \; .
\label{MMCgamma}
\end{equation}
Using $\gamma_{\rm \scriptscriptstyle MMCS}$ in the Callan-Symanzik equation 
(\ref{CSeqn}) not only explains the leading logarithms 
(\ref{MMCWeyl}-\ref{MMCNewton}), but also permits a full resummation,
\begin{eqnarray}
C_{0i0j}(t,\vec{x}) &\!\!\! \longrightarrow \!\!\!& C^{\rm tree}_{0i0j}(t,\vec{x})
\times \Bigl[ a(t)\Bigr]^{-\frac{3 \kappa^2 H^2}{160 \pi^2}} \; , \qquad 
\label{MMCWeylresum} \\
\Psi(t,r) &\!\!\! \longrightarrow \!\!\!& \tfrac{G M}{a(r) r} \times
\Bigl[ a(t) r H\Bigr]^{-\frac{3 \kappa^2 H^2}{160 \pi^2}} \; . \qquad
\label{MMCNewtonresum}
\end{eqnarray}
For massless, conformally coupled scalars one has $c_1 = 0$ and $c_2$ 
equal to \cite{Foraci:2024cwi},
\begin{equation}
c_2 = \tfrac{\mu^{D-4} \Gamma(\frac{D}2)}{2^8 \pi^{\frac{D}2}}
\tfrac{2}{(D+1) (D-1) (D-3)^2 (D-4) } \qquad \Longrightarrow \qquad
\gamma_{\rm \scriptscriptstyle MCCS} = -\tfrac{\kappa^2 H^2}{960 \pi^2} \; .
\label{MCCc2}
\end{equation}
This not only explains the leading logarithms at 1-loop 
(\ref{MCCWeyl}-\ref{MCCNewton}), and again implies the resumed results,
\begin{eqnarray}
C_{0i0j}(t,\vec{x}) &\!\!\! \longrightarrow \!\!\!& C^{\rm tree}_{0i0j}(t,\vec{x})
\times \Bigl[ a(t)\Bigr]^{\frac{\kappa^2 H^2}{480 \pi^2}} \; , \qquad 
\label{MCCWeylresum} \\
\Psi(t,r) &\!\!\! \longrightarrow \!\!\!& \tfrac{G M}{a(r) r} \times
\Bigl[ a(t) r H\Bigr]^{\frac{\kappa^2 H^2}{480 \pi^2}} \; . \qquad
\label{MCCNewtonresum}
\end{eqnarray}
A loop of Dirac fermions has $\gamma_{\rm \scriptscriptstyle Dirac} = 6 \times
\gamma_{\rm \scriptscriptstyle MMCS}$ \cite{Foraci:2024cwi} and a loop of
photons gives $\gamma_{\rm \scriptscriptstyle EM} = 12 \times
\gamma_{\rm \scriptscriptstyle MMCS}$ \cite{Foraci:2024vng}.

\section{Conclusions}

Alexei Starobinsky possessed a towering intellect which allowed him to
see deeply hidden truths. One of these is that a simple stochastic
formalism can describe many of the secular loop corrections one finds 
in quantum field theory on inflationary backgrounds. This is quite
surprising from the context of quantum field theory because the 
fields which obey Starobinsky's first order Langevin equation are 
ultraviolet finite, commuting variables, quite unlike the original
quantum fields which obey second order equations, harbor ultraviolet
divergences, and do not commute on or within the light-cone.

The message of this article is that Starobinsky's stochastic formalism 
is neither wrong, as many QFT experts believed, nor does it supplant 
quantum field theory, as some cosmologists believed. What it represents
is the leading secular corrections in quantum field theories of
undifferentiated ``active'' fields which experience inflationary 
particle production (\ref{occupation}). It does not directly apply to 
``passive'' fields, which do not experience inflationary particle 
production, or to differentiated active fields. Those systems make 
nonsecular, but still nonzero, contributions which come as much from 
the ultraviolet as from the infrared, and from the full mode function. 
The correct way to deal with these systems is by integrating out the 
problematic passive or differentiated active fields in the presence of 
a constant active field background. For passive fields this typically 
results in a standard effective potential induced by a field-dependent 
mass. For differentiated active fields one encounters new sorts of 
effective potentials induced by field-dependent field strengths and/or 
field-dependent Hubble parameters. 

Finally, it is best avoid dogmatic adherence to statistical mechanics,
and to check against explicit computations whenever possible. Although 
Starobinsky's formalism does have a statistical mechanical interpretation,
its derivation rests on quantum field theory, and what it does is to 
predict a certain class of secular corrections. One must renormalize in
the presence of either passive fields or differentiated active fields. 
And differentiated active fields produce new types of secular corrections 
from renormalization (\ref{renorm}), which cannot be represented by any
stochastic formalism. This second source of secular corrections was only
discovered by explicit 1-loop and 2-loop computations. It is possible 
that further surprises await us, which Alexei would have loved.

\newpage

\centerline{\bf Acknowledgements}

It is a pleasure to acknowledge collaboration, conversation and 
correspondence on these subjects with A. Foraci, D. Glavan, S. P. Miao, 
T. Prokopec, N. C. Tsamis and B. Yesilyurt. This work was partially 
supported by NSF grant PHY-2207514 and by the Institute for Fundamental 
Theory at the University of Florida.

\end{document}